\documentclass[twocolumn,showpacs,pra]{revtex4}%
\usepackage{amsfonts}
\usepackage{amsmath}
\usepackage{amssymb}
\usepackage{graphicx}%
\providecommand{\U}[1]{\protect\rule{.1in}{.1in}}

\begin{document}
\title{Lyapunov exponent for the laser speckle potential: a weak disorder expansion}
\author{Evgeni Gurevich}
\author{Oded Kenneth}
\affiliation{Department of Physics, Technion, Israel Institute of Technology, Haifa 32000, Israel}

\begin{abstract}
Anderson localization of matter waves was recently observed with cold atoms in
a weak 1D disorder realized with laser speckle potential
\cite{Aspect-1DBECexper-08}. The latter is special in that it does not have
spatial frequency components above certain cutoff $q_{c}$. As a result, the
Lyapunov exponent (LE), or inverse localization length, vanishes in Born
approximation for particle wavevector $k>\frac{1}{2}q_{c}$, and higher orders
become essential. These terms, up to the order four, are calculated
analytically and compared with numerical simulations. For very weak disorder,
LE exhibits a sharp drop at $k$ $=\frac{1}{2}q_{c}$. For moderate disorder (a)
the drop is less dramatic than expected from the fourth order approximation
and (b) LE becomes very sensitive to the sign of the disorder skewness (which
can be controlled in cold atom experiments). Both observations are related to
the strongly non-Gaussian character of the speckle intensity.

\end{abstract}
\startpage{1}
\endpage{20}
\maketitle

Technological progress in experiments with ultracold atoms provides an
extraordinary level of control, allowing investigation of various quantum
phenomena \cite{BEC-Rev}. One such phenomenon is Anderson localization (AL)
\cite{Anderson-58} of matter waves. Formulated originally to explain the
absence of spreading of quantum-mechanical wave function in a disordered
potential, it was recognized later as a common signature of wave propagation
in random media, where waves may become exponentially localized because of the
destructive interference between its multiple scattered components. Some
indications of AL were observed with light \cite{Light-Loc}, microwaves
\cite{Micro-W-Loc} and ultrasound \cite{Sound-Loc}, while in electronic
systems it is hindered by finite temperature dephasing and interactions.

Cold atoms offer a unique possibility to study matter wave localization in
conditions of very low temperature and tunable interactions (not available in
solids), absence of absorption and controllable disorder. The latter may be
introduced by different techniques, one of which is using static laser
speckle, whereas potential felt by atoms is proportional to the speckle
intensity with the sign of the detuning from the atomic transition
\cite{Aspect-Speckle-06}. Laser speckle, generated by passing expanded laser
beam through diffusive plates, are special in that they have (i) exponential,
i.e. strongly non-Gaussian, intensity distribution and (ii) finite support of
their power spectrum \cite{Goodman-84}. As discussed below, these properties
have strong effect on the localization properties in 1D systems.

Recently, AL was observed in Bose-Einstein condensates expanding in weak 1D
disorder realized with laser speckle \cite{Aspect-1DBECexper-08}. A simplified
model for such experiments considers expansion of the condensate released from
a harmonic trap as occurring in two stages \cite{Aspect-1DBECtheor-07}%
,\cite{Shapiro-07}: (1) explosive conversion of the interaction energy into
kinetic energy, in which weak disorder may be neglected, followed by (2)
expansion of the non-interacting gas in the disordered potential. The first
stage ends up with a certain momentum distribution of the atomic cloud
\cite{BEC-explosion}, which provides an initial condition for the second
stage. Thus, the problem reduces to basically a single-particle localization,
characterized in 1D by Lyapunov exponent (LE), or inverse localization length
\cite{LGP-Introduction}. The latter, in a weak 1D speckle disorder with a
spatial frequency cutoff $q_{c}$, has some peculiarities, because in 1D (as
opposed to higher dimensions), elastic scattering is a "binary" process:
either the particle wavevector $k$ remains unchanged (forward scattering), or
it reverses its sign and changes by the amount of $2k$ (backscattering). As a
result, backscattering amplitude (in a single potential realization) vanishes
in the Born approximation for $2k>q_{c}$, while in higher approximations, of
order $n$, it vanishes for $2k>nq_{c}$. Correspondingly, in \emph{weak}
speckle disorder, LE is expected to exhibit a series of cascading drops at
$k=\frac{n}{2}q_{c}$, so called "effective mobility edges". Thus, in order to
assess the localization properties for $2k>q_{c}$, one should know LE beyond
the Born approximation, which is a subject of this paper.

LE in 1D disorder was studied extensively for a variety of continuous and
discrete models, which usually dealt with an uncorrelated disorder or a
specific type of correlation (see Ref.\cite{LGP-Introduction} and reference
herein). For arbitrary correlation and in the weak disorder Born
approximation, LE is known to be proportional to the disorder power spectrum
\cite{LGP-Introduction},\cite{Israilev-99}. In the recent years, correlations
in 1D disorder attracted much attention because they can induce unusual
localization properties, such as existence of extended states
\cite{Extend-Stats}, deviation from the single parameter scaling (SPS)
\cite{SPS} and appearance of the "effective mobility edge" when LE vanishes in
Born approximation \cite{Israilev-99},\cite{Tessieri-02}. The latter means
that exact localization length would exceed system size for sufficiently weak
disorder, while higher orders are required to specify this regime
quantitatively \cite{Tessieri-02}.

In this paper we report a systematic weak disorder expansion for LE for two
orders beyond Born approximation, which is then applied to the laser speckle
disorder \cite{Aspect-private}. The analytical study is verified by numerical
simulations, which also allow examination of regimes beyond weak disorder. The
author of Ref.\cite{Tessieri-02} performed a weak disorder expansion for a
"generalized" LE. Let us stress that, although the generalized LE is easier to
compute, it is the standard LE, studied here, that is of prime interest in the
localization problem. While these two quantities coincide in the lowest order
\cite{Tessieri-02}, they differ in higher orders (see below). More generally,
the equality holds under the assumption of SPS, and the differences found show
how the latter is affected by disorder correlations.

We consider continuous one-dimensional model%
\begin{equation}
\frac{d^{2}\psi}{dx^{2}}+k^{2}(1+\eta(x))\psi=0 \label{eqn1}%
\end{equation}
where $\eta(x)$ is the dimensionless disordered potential (for a quantum
particle, $\eta(x)=-2mV(x)/\hbar^{2}k^{2}$ is the ratio of the potential
$V(x)$ to the particle energy). We assume zero mean finite-range correlated
disorder, specified by its (joint) cumulants ($n=2,3,\ldots$)%
\begin{equation}
\kappa_{n}\left(  x_{1},\ldots,x_{n-1}\right)  =\left(  \frac{g}{2R_{c}%
}\right)  ^{\frac{n}{2}}\Gamma_{n}\left(  \frac{x_{1}}{R_{c}},\ldots
,\frac{x_{n-1}}{R_{c}}\right)  , \label{Eq. Nth cumulant notation}%
\end{equation}
where $R_{c}$ is the correlation scale, the dimensionless functions
$\Gamma_{n}\left(  x_{1},\ldots,x_{n-1}\right)  $ decay on the scale of unity
and $g$ is noise intensity. One chooses $\int_{0}^{+\infty}\Gamma_{2}(x)dx=1$,
so that the limit $R_{c}\rightarrow0$ yields Gaussian white noise with the
two-point correlation $\left\langle \eta(x)\eta(x^{\prime})\right\rangle
=g\delta(x-x^{\prime})$.

LE is defined as
\begin{equation}
\lambda=\underset{x\rightarrow\infty}{\lim}\left(  2x\right)  ^{-1}%
\left\langle \ln(k^{2}\left\vert \psi\left(  x\right)  \right\vert
^{2}+\left\vert \psi^{\prime}\left(  x\right)  \right\vert ^{2})\right\rangle
, \label{LE definition}%
\end{equation}
where $\left\langle ..\right\rangle $ denotes the disorder average, while the
\emph{generalized} LE considered in Ref.\cite{Tessieri-02} is given by
$\underset{x\rightarrow\infty}{\lim}\frac{1}{4x}\ln\langle k^{2}\left\vert
\psi\left(  x\right)  \right\vert ^{2}+\left\vert \psi^{\prime}\left(
x\right)  \right\vert ^{2}\rangle$. We calculate the LE (\ref{LE definition})
using the phase formalism relation \cite{LGP-Introduction}%
\begin{equation}
\lambda=\int zP_{st}\left(  z\right)  dz,
\end{equation}
where $z=\psi^{\prime}/\psi$ and $P_{st}\left(  z\right)  $ is the stationary
(i.e. the $x\rightarrow\infty$ limit) distribution of $z$. Introducing phase
$\theta$, defined by $\left.  z=-k\tan\left(  \theta/2\right)  \right.  $ and
obeying "evolution" equation%
\begin{equation}
\frac{\partial\theta}{\partial x}=2k+2k\cos^{2}(\theta/2)\eta(x),
\label{eq. Evol Eq. for phase}%
\end{equation}
LE is expressed in terms of the stationary distribution of the phase as%
\begin{equation}
\lambda=-k\int_{-\pi}^{\pi}\tan\left(  \frac{\theta}{2}\right)  P_{st}\left(
\theta\right)  d\theta. \label{LE via P(theta)}%
\end{equation}
Then, calculating the weak disorder expansion for $P_{st}\left(
\theta\right)  $ and substituting it into (\ref{LE via P(theta)}), yields the
required expansion for the LE $\lambda$. In the case in which $\eta(x)$ in
(\ref{eq. Evol Eq. for phase}) is a $\delta$-correlated process, one can use
standard technique \cite{Risken} to write down a Fokker-Plank equation for
$P\left(  \theta;x\right)  $. When $\eta(x)$ is a correlated process, the
method of "ordered cumulants" \cite{V-Kampen-book} can be used to obtain an
approximate master equation for $P(\theta;x)$, given by a perturbative
expansion in powers of $\beta\gamma^{1/2}$, where the dimensionless parameters%
\begin{equation}
\gamma=2kR_{c}\text{ \ and \ }\beta=\sqrt{gk/8},
\end{equation}
describe disorder correlation scale and strength respectively. Specific
application of the method to the present problem involves many technical
details, which will be given elsewhere \cite{EGur}. Here we only outline the
main steps of the derivation. Assuming small $\beta$, but arbitrary $\gamma$,
the following formal expansion of the master equation for $P\left(
\theta;x\right)  $ is obtained%
\begin{equation}
\frac{\partial}{\partial x}P(\theta;x)=\left[  A_{0}+2\sum\nolimits_{n=2}%
^{\infty}\beta^{n}K_{n}\right]  P(\theta;x). \label{eq. Master Eq. Gen}%
\end{equation}
$K_{n}$ are differential operators, whose definition involves integration over
various ordered products of the operator $\tilde{A}_{1}(x)\equiv e^{xA_{0}%
}A_{1}e^{-xA_{0}}=-2k\partial_{\theta}\cos^{2}\left(  \theta-kx\right)  $,
weighted with combinations of the cumulant functions $\Gamma_{m\leq n}$
(\ref{Eq. Nth cumulant notation}). Here operators $A_{0}=-2k\partial_{\theta}$
and $A_{1}=-2k\partial_{\theta}\cos^{2}(\theta/2)$ are related respectively to
the deterministic and the stochastic terms on the right side of
(\ref{eq. Evol Eq. for phase}).

Next, in the stationary limit, the master equation for $P(\theta;x)$ reduces
to an ordinary differential equation for $P_{st}\left(  \theta\right)
=\lim\limits_{x\rightarrow\infty}P(\theta;x)$ and one looks for a perturbative
solution%
\begin{equation}
\label{Pst-exp}P_{st}\left(  \theta\right)  =\sum\beta^{n}P_{n}\left(
\theta\right)  .
\end{equation}
Let us note that this "double-stage" perturbation approach is inconsistent in
some cases \cite{Hangi-95},\cite{V-Kampen-89}, such as when $P_{st}(\theta)$
is singular in the limit $\beta\rightarrow0$. This, however, does not occur in
our case, since $P_{st}(\theta)$ becomes uniform for $\beta\rightarrow0$.
Thus, substitution of the solution (\ref{Pst-exp}) into (\ref{LE via P(theta)}%
) yields the required LE expansion%
\begin{equation}
\lambda=k\sum\nolimits_{n\geq2}\beta^{n}\lambda_{n}\left(  \gamma\right)  .
\label{LE expansion general}%
\end{equation}
The first four coefficients are given by%
\begin{equation}
\lambda_{2}=c_{o},\lambda_{3}=c_{6},\lambda_{4}=c_{0}\left(  c_{1}%
-c_{2}\right)  +c_{3}c_{1}+2c_{4}+c_{5}, \label{Lambdas(c)}%
\end{equation}
where $c_{i}$ are the following functions of $\gamma$($=2kR_{c}$):%
\begin{align}
c_{0}  &  =\int\nolimits_{0}^{\infty}ds\Gamma_{2}\left(  s\right)  \cos\gamma
s=\frac{\tilde{\Gamma}_{2}\left(  \gamma\right)  }{2},\text{\quad}%
c_{3}=-\gamma\frac{\partial c_{0}}{\partial\gamma},\\
c_{1}  &  =\int\nolimits_{0}^{\infty}ds\Gamma_{2}\left(  s\right)  \sin\gamma
s=\int_{-\infty}^{+\infty}\frac{dq}{2\pi}\frac{\tilde{\Gamma}_{2}\left(
q\right)  }{\gamma-q},\text{\ }c_{2}=\gamma\frac{\partial c_{1}}%
{\partial\gamma},\nonumber
\end{align}%
\begin{align}
c_{4}  &  =\int\nolimits_{0}^{\infty}ds_{1}\int\nolimits_{0}^{s_{1}}%
ds_{2}\Gamma_{2}\left(  s_{1}\right)  \Gamma_{2}\left(  s_{2}\right)
\times\nonumber\\
&  \text{\quad}\times\left[  2\sin\left(  \gamma s_{1}\right)  -\gamma\left(
s_{1}-s_{2}\right)  \cos\left(  \gamma s_{1}\right)  \right] \\
&  =\int\frac{dq}{2\pi}\tilde{\Gamma}_{2}\left(  q\right)  \left[
\frac{\tilde{\Gamma}_{2}\left(  \gamma-q\right)  }{q}+\gamma\frac
{\tilde{\Gamma}_{2}\left(  \gamma-q\right)  -\tilde{\Gamma}_{2}\left(
\gamma\right)  }{2q^{2}}\right]  ,\nonumber
\end{align}%
\begin{align}
c_{5}  &  =-\gamma\int\nolimits_{0}^{\infty}ds_{1}\int\nolimits_{s_{1}%
}^{\infty}ds_{2}\int\nolimits_{s_{2}}^{\infty}ds_{3}\Gamma_{4}\left(
s_{1},s_{2},s_{3}\right)  \times\nonumber\\
&  \text{\quad}\times\left(  2\cos\left(  \gamma s_{3}\right)  +\cos\left(
\gamma s_{1}-\gamma s_{2}-\gamma s_{3}\right)  \right)  ,
\end{align}%
\begin{align}
\text{ }c_{6}  &  =-\sqrt{2\gamma}\int\nolimits_{0}^{\infty}ds_{1}%
\int\nolimits_{s_{1}}^{\infty}ds_{2}\Gamma_{3}\left(  s_{1},s_{2}\right)
\sin\gamma s_{2}\nonumber\\
&  =\frac{\sqrt{2\gamma}}{4\pi}\int\frac{dq}{q}\left[  \tilde{\Gamma}%
_{3}\left(  q,\gamma\right)  -\tilde{\Gamma}_{3}\left(  q,\gamma-q\right)
\right]  ,
\end{align}
and $\tilde{\Gamma}_{n}\left(  q_{1},\ldots,q_{n-1}\right)  $ \ is a Fourier
transform of $\Gamma_{n}\left(  x_{1},\ldots,x_{n-1}\right)  $. The
coefficients $c_{i}$ for $i=5,6$ may be tagged "non-Gaussian", since they
depend on the higher (non-Gaussian) cumulants $\Gamma_{3}$ and $\Gamma_{4}$
only, and vanish for Gaussian disorder. Note that $c_{i=0,3,4,6}$ would vanish
for $\gamma$ above certain threshold, if $\tilde{\Gamma}_{2}\left(  q\right)
$ and $\tilde{\Gamma}_{3}\left(  q_{1},q_{2}\right)  $ have finite support
(same applies to $c_{5}$, whose expression in terms of $\tilde{\Gamma}_{4}$ is
not shown). Besides, if $\tilde{\Gamma}_{2}\left(  q\right)  $ or its
derivative are discontinuous at some point $q_{c}$, then $c_{4}\left(
\gamma\right)  $ would diverge at $\gamma=q_{c}$ and the perturbation theory
would break down for this $\gamma$ (see below).

As expected, $\lambda_{2}$, the lowest order coefficient in
(\ref{LE expansion general}), coincides with earlier results
\cite{LGP-Introduction}. Comparing our results for the standard LE to those of
Ref.\cite{Tessieri-02} for the generalized LE (even orders only), one finds
that the two coincide in the second and vary in the fourth order, though the
difference exists only for non-Gaussian disorder. Namely, the corresponding
non-Gaussian terms $c_{5}\left(  \gamma\right)  $ in $\lambda_{4}$ have
different expressions, which at small $\gamma$ scale differently with $\gamma$
and have opposite signs. Thus, the SPS relations (see e.g. Ref.\cite{SPS})
hold up to the order four in Gaussian, while are broken in non-Gaussian disorder.

So far our results are quite general and pertain to an arbitrary random
potential, the only condition being sufficiently fast decay of the disorder
cumulants \cite{V-Kampen-book}. Now we specialize to the case of a laser
speckle potentials, produced by transmitting laser beam through a diffuser
with a $\emph{rectangular}$ aperture \cite{Goodman-84}%
,\cite{Aspect-Speckle-06}. Its intensity pair correlation function is
$\Gamma_{2}\left(  x\right)  =\frac{2}{\pi}\frac{\sin^{2}x}{x^{2}}$, whose
Fourier transform $\tilde{\Gamma}_{2}\left(  q\right)  $ vanishes for
$\left\vert q\right\vert \geq2$ ($\tilde{\Gamma}_{2}\left(  q\right)  $ is
related to the shape of the optical aperture, therefore for arbitrary but
finite aperture it would always have finite support \cite{Goodman-84}).
Assuming that speckle $\emph{field}$ (as opposed to intensity) is a complex
Gaussian variable, which is true for sufficiently large diffuser
\cite{Goodman-84}, one concludes that any $\Gamma_{n}$ is expressed solely in
terms of "irreducible" products of the two-point \emph{field} correlators
$\tilde{w}\left(  x_{i}-x_{j}\right)  $. For example, the $3^{rd}$ cumulant is
$\Gamma_{3}\left(  x_{1},x_{2}\right)  =-2\varepsilon\left(  \frac{2}{\pi
}\right)  ^{3/2}\tilde{w}\left(  x_{1}\right)  \tilde{w}\left(  x_{1}%
-x_{2}\right)  \tilde{w}\left(  x_{2}\right)  $, where $\tilde{w}\left(
x\right)  =\frac{\sin x}{x}$ is related to the Fourier transform of the
"rectangle" function and $\varepsilon=\pm1$ is the sign of the disorder
distribution skewness, depending on either "blue" or "red" laser detuning from
the atomic transition.

\begin{figure}
[tb]
\begin{center}
\includegraphics[
height=2.4018in,
width=3.4728in
]%
{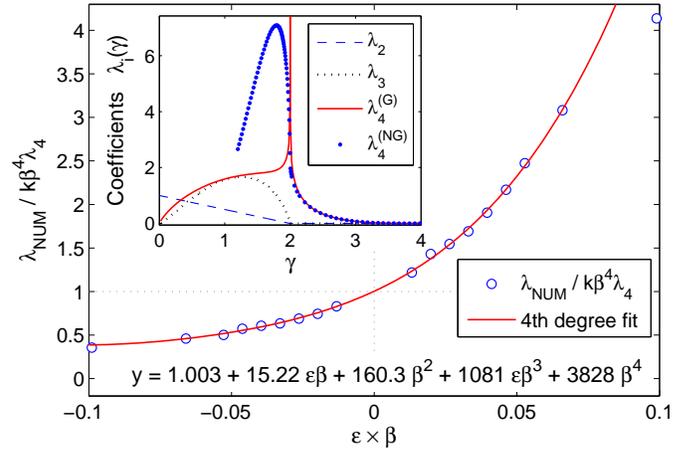}%
\caption{(color online) LE calculated at fixed $\gamma=2.2$ as a function of
disorder strength and "sign" $\varepsilon\beta$. Circles represent numerical
LE devided by $k\beta^{4}\lambda_{4}$. Line and equation show the $4^{th}$
degree fit over all the points except the left- and the right-most ones.
Insert: expansion coefficients $\lambda_{i}\left(  \gamma\right)  $. }%
\label{Fig. LE betta fit}%
\end{center}
\end{figure}

Substituting the explicit expressions for $\Gamma_{n}$ into the definitions of
the coefficients $c_{i}$, one obtains the following LE expansion coefficients
for the speckle disorder:%
\begin{align}
\lambda_{2}  &  =\frac{2-\gamma}{2}\chi\left(  2-\gamma\right)  ,\quad
\lambda_{4}=\left(  \lambda_{4}^{G}+\lambda_{4}^{NG}\right)  \chi\left(
4-\gamma\right)  ,\nonumber\\
\lambda_{3}  &  =2\varepsilon\sqrt{\frac{\gamma}{\pi}}\left[  \left(
\gamma-2\right)  \ln\frac{2-\gamma}{2}-\gamma\ln\frac{\gamma}{2}\right]
\chi\left(  2-\gamma\right)  , \label{Lambdas}%
\end{align}
where $\lambda_{4}^{G}\equiv c_{0}\left(  c_{1}-c_{2}\right)  +c_{3}%
c_{1}+2c_{4}$ and $\lambda_{4}^{NG}\equiv c_{5}$ denote the "Gaussian" and the
non-Gaussian parts of $\lambda_{4}$ and $\chi\left(  x\right)  $ is the
Heaviside step function. For $\gamma\leq2$%
\begin{align}
\lambda_{4}^{G}  &  =\frac{1}{2\pi}\left[  4\gamma-3\gamma^{2}+\left(
\frac{\gamma^{2}}{2}+2\right)  \ln\frac{\gamma+2}{2}+\right. \nonumber\\
&  \quad\left.  +\allowbreak\left(  \frac{3\gamma^{2}}{2}-10\right)  \ln
\frac{2-\gamma}{2}+\gamma\left(  \gamma-4\right)  \ln\frac{\gamma}{2}\right]
,\\
\lambda_{4}^{NG}  &  \approx\frac{\gamma\pi}{3}+\frac{\gamma}{\pi}\left(
2-\gamma\right)  \ln\left(  2-\gamma\right)  \left(  \frac{11}{2}\ln
\frac{2-\gamma}{4e^{2}}-\ln2\right) \nonumber
\end{align}
while for $2<\gamma<4$%
\begin{align}
\lambda_{4}^{G}  &  =\frac{1}{2\pi}\left[  \gamma^{2}-4\gamma-\left(
\gamma^{2}-4\gamma+8\right)  \ln\frac{\gamma-2}{2}\right]  ,\\
\lambda_{4}^{NG}  &  =\frac{\gamma^{2}}{\pi}\left[  2\operatorname{Li}%
_{2}\frac{2}{\gamma}-\frac{\pi^{2}}{6}+\ln^{2}\frac{\gamma}{2}\right]
-\frac{\gamma\left(  \gamma-2\right)  }{\pi}\ln^{2}\frac{\gamma-2}%
{2},\nonumber
\end{align}
where $\operatorname{Li}_{2}\left(  z\right)  $ is the dilogarithm and
$\lambda_{4}^{NG}$ for $\gamma<2$ was calculated approximately, assuming
$\left(  2-\gamma\right)  \ll1$ and neglecting $\mathcal{O}\left(
2-\gamma\right)  $ terms (exact calculation is too lengthy to carry out, while
$\lambda_{4}^{NG}$ is important only close to or above $\gamma=2$). The
coefficients $\lambda_{i}\left(  \gamma\right)  $ are plotted in the insert of
Fig. \ref{Fig. LE betta fit}. As already noticed \cite{Aspect-Speckle-06},
$\lambda_{2}$ vanishes for $\gamma=2kR_{c}>2$. Then, as can be expected from
the general perturbation theory for scattering, $\lambda_{3}$ and $\lambda
_{4}$ vanish for $\gamma>2$ and $\gamma>4$ respectively. Thus, for
$2<\gamma<4$, LE switches from quadratic to quartic leading dependence on
$\beta$ and, for weak disorder, undergoes a steep decrease at $\gamma
=2kR_{c}=2$. Note that for speckle potential our perturbation theory breaks
down at $\gamma=2$, where $\lambda_{4}^{G}$ has a logarithmic divergence. As
explained above, this is because $\tilde{\Gamma}_{2}\left(  q\right)  =$
$\left(  2-\left\vert q\right\vert \right)  \chi\left(  2-\left\vert
q\right\vert \right)  $ has a discontinuous derivative at $q=2$.

\begin{figure}
[ptb]
\begin{center}
\includegraphics[
height=2.1976in,
width=3.3981in
]%
{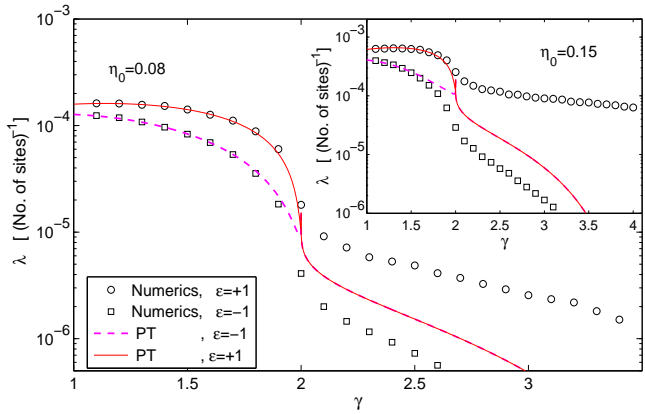}%
\caption{(color online) Lyapunov exponent $\lambda$ for\thinspace
\thinspace$\eta_{0}=\allowbreak0.08$ (main panel) and $\eta_{0}=\allowbreak
0.15$ (insert) as a function of the dimensionless disorder correlation
parameter $\gamma=2kR_{c}$. Circles and squares show numerical data for
positive ($\varepsilon=+1$) and negative ($\varepsilon=-1$) skewness of
disorder, while solid and dashed lines give corresponding $4^{th}$ order
expansion. Curve's spike at $\gamma=2$ reflects the logarithmic divergence of
$\lambda_{4}^{G}$.}%
\label{Fig. LE STD.004+.0075}%
\end{center}
\end{figure}

Our analytical study was supplemented with numerical simulations of the
discrete tight-binding model near the energy band edge, where it is a good
approximation to continuous problem. LE was computed using the transmission
matrix formalism \cite{Liu-86}. LE, as a function of $\gamma$ at fixed
$\eta_{0}\equiv\sqrt{\left\langle \eta^{2}\right\rangle }$, the ratio between
the disorder standard deviation and the particle energy, is shown in Fig.
\ref{Fig. LE STD.004+.0075} for $\eta_{0}=0.08$ and $\eta_{0}=0.15$ (the
disorder strength parameter $\beta$, rewritten as $\beta=\frac{1}{4}\sqrt
{\pi\gamma}\eta_{0}$, changes along with $\gamma$ in this parametrization).
LE, for each $\eta_{0}$ computed for both signs of $\varepsilon$, is
significantly larger for $\varepsilon=+1$, which demonstrates strong effect of
the non-Gaussian character of the speckle disorder.%
While agreement between the analytical and the numerical results is acceptable
for $\gamma<2$, it appears to be very poor for $\gamma>2$, which is because
the disorder is not weak enough. To clarify this point, numerical LE
($\lambda^{\left(  Num\right)  }$), computed for different values of $\beta$
and $\varepsilon$ at fixed $\gamma=2.2$ and divided by the analytical
$\lambda=k\beta^{4}\lambda_{4}$, was fitted to a $4^{th}$ degree polynomial of
$\varepsilon\beta$, as appears in Fig. \ref{Fig. LE betta fit} together with
the fit equation (recall that for $\gamma>2$ the perturbative expansion
(\ref{LE expansion general}) for $\lambda$ starts with the fourth order term).
The free constant of the fit is $\lambda_{4}^{\left(  Num\right)  }%
/\lambda_{4}$, and its value $1.003$ indicates excellent agreement between the
perturbation theory and numerics. On the other hand, the fit equation in Fig.
\ref{Fig. LE betta fit} shows that the $4^{th}$ order approximation is
acceptable only for very weak disorder ($\beta\ll0.1$). Repeating similar test
for $\gamma\approx2$, we conclude that the logarithmic peak of $\lambda
_{4}^{G}$ (Fig. \ref{Fig. LE betta fit}, insert) is reproduced in the numerics
up to the smearing effect of the finite system size $L$ (this can be accounted
for by smoothing $\lambda_{i}\left(  \gamma\right)  $ over $\Delta\gamma
\sim\frac{R_{c}}{L}$).

Note the very rapid growth of the higher order coefficients of the fit in Fig.
\ref{Fig. LE betta fit}, corresponding to higher orders in LE expansion. As
suggested by the structure of the perturbation theory, it is related to the
non-Gaussian character of the exponential distribution of the speckle
intensity, whose cumulants grow factorially. This explains both the failure of
the $4^{th}$ order approximation and the strong dependence of LE on the sign
of the disorder skewness $\varepsilon$ found at \emph{moderately} weak
disorder (Fig. \ref{Fig. LE STD.004+.0075}). The perturbative expansion up to
order $\beta^{4}$ accounts for the effect of $\varepsilon$ only for $\gamma
<2$, since its only odd term $\lambda_{3}$ vanishes for $\gamma>2$. Therefore,
analytical curves for $\varepsilon=+1$ and $\varepsilon=-1$ coincide for
$\gamma>2$ (Fig. \ref{Fig. LE STD.004+.0075}).

Finally, we address persistence of the effective mobility edge at $\gamma=2$
as a function of the disorder strength. This question is important for the
recent experiments on the BEC expansion \cite{Aspect-1DBECexper-08} (with
$\varepsilon=+1$), because, if distribution of the atomic $k$'s in the
"exploded" condensate stretches beyond the speckle frequency cutoff $2/R_{c}$,
then the deepness of the LE drop at $\gamma=2$ becomes crucial for predicting
algebraic versus exponential decay of density profiles
\cite{Aspect-1DBECtheor-07}. Our computations in Fig.
\ref{Fig. LE STD.004+.0075} show that this drop depends strongly on the
disorder strength and becomes of effectively one order of magnitude or less
for $\eta_{0}\gtrsim0.1$ (relevant to the experimental regimes
\cite{Aspect-1DBECexper-08}). In addition, note the difference in LE for
"blue" ($\varepsilon=+1$) and "red" ($\varepsilon=-1$) detuning: for negative
detuning LE is smaller, but its relative variation across $\gamma=2$ is
larger. (Though our parametrization is not natural for this experiment - we
fixed $k$ with disorder amplitude and varied $R_{c}$, instead of varying $k$
with the rest fixed, - the conclusions remain valid, since mainly the behavior
near $\gamma=2$ is concerned.)

In conclusion, we derived general expansion for the Lyapunov exponent (LE) in
1D correlated disorder two orders beyond the Born approximation. Comparing it
with that for the generalized LE \cite{Tessieri-02} shows that single
parameter scaling is broken perturbatively in the fourth order for
\emph{non-Gaussian} disorders (with finite moments). Applying this expansion
to speckle disorder with Fourier spectrum cutoff $q_{c}=\frac{2}{R_{c}}$, we
find that the leading order dependence of LE on the disorder strength crosses
from quadratic for $kR_{c}<1$ to quartic for $1<kR_{c}<2$, as expected from
the standard QM perturbation theory. For very weak disorder, this results in
large and steep drop of LE across $kR_{c}=1$. For larger but still weak
disorder, this drop moderates, while LE becomes very sensitive to the skewness
of the disorder distribution (for $kR_{c}\gtrsim1$). This is because of
strongly non-Gaussian distribution of speckle intensity, whose cumulants grow
factorially fast. Physically, this reflects non-perturbative contribution of
rare but large potential peaks (dips) of the typical width $R_{c}$ to the
scattering process.

\begin{acknowledgments}
E.G. thanks B. Shapiro for suggesting the problem and for his guidance. We
wish to thank K. Mallick for valuable discussions and are grateful to A.
Aspect for an illuminating discussion and for informing us about the work on
correlated random potentials carried on in his and collaborating groups
\cite{Aspect-private}. This research was supported by ISF grant.
\end{acknowledgments}


\begin{thebibliography}{99}                                                                                               %


\bibitem {BEC-Rev}For a recent review, see I. Bloch \textit{et al.},
\textit{Rev. Mod. Phys} \textbf{80,} 885 (2008)

\bibitem {Anderson-58}P. W. Anderson, Phys. Rev. \textbf{109,} 1492 (1958).

\bibitem {Light-Loc}D. S. Wiersma \textit{et al.}, \textit{Nature}
\textbf{390}, 671673 (1997); T. Schwartz \textit{et al.}, \textit{Nature}
\textbf{446}, 5255 (2007); Y. Lahini \textit{et al.}, Phys. Rev. Lett.
\textbf{100}, 013906 (2008).

\bibitem {Micro-W-Loc}Chabanov \textit{et al.}, \textit{Nature} \textbf{404},
850853 (2000).

\bibitem {Sound-Loc}H. Hu \textit{et al.}, \textit{Nature Physics} \textbf{4},
945 (2008).

\bibitem {Aspect-Speckle-06}D. Cl\'{e}ment \textit{et al.}, \textit{New J.
Phys} \textbf{8,} 165 (2006); L. Sanchez-Palencia \textit{et al.}, \textit{New
J. Phys} \textbf{10}, 045019 (2008).

\bibitem {Goodman-84}J.W. Goodman, "\textit{Statistical properties of Laser
Speckle Patterns}" in "\textit{Laser Speckle and Related Phenomena}", edited
by J.C. Dainty, 2nd Ed., Springer-Verlag 1984.

\bibitem {Aspect-1DBECexper-08}J. Billy \textit{et al.,} \textit{Nature}
\textbf{453}, 891 (2008).

\bibitem {Aspect-1DBECtheor-07}L. Sanchez-Palencia \textit{et al.},
\textit{Phys. Rev. Lett.} \textbf{98}, 210401 (2007).

\bibitem {Shapiro-07}B. Shapiro, \textit{Phys. Rev. Lett.} \textbf{99}, 060602 (2007).

\bibitem {BEC-explosion}Yu. Kagan \textit{et al.}, \textit{Phys. Rev.} A
\textbf{54}, R1753 (1996); Y. Castin and R. Dum,\textit{ Phys. Rev. Lett.}
\textbf{77}, 5315 (1996).

\bibitem {LGP-Introduction}I.M. Lifshitz \textit{et al.}, \textit{Introduction
to the Theory of Disordered Systems} (Wiley, New York, 1988).

\bibitem {Extend-Stats}D. H. Dunlap \textit{et al.}, \textit{Phys. Rev. Lett.}
\textbf{65}, 88 (1990); F.A.B.F. de Moura and M. L. Lyra, \textit{Phys. Rev.
Lett.} \textbf{81}, 3735 (1998); P. Carpena \textit{et al.}, \textit{Nature}
\textbf{418}, 955 (2002); A. M. Garcia-Garcia and E. Cuevas, arXiv:0808.3757v1.

\bibitem {SPS}M. Titov and H. Schomerus, \textit{Phys. Rev. Lett.}
\textbf{95}, 126602 (2005).

\bibitem {Israilev-99}F.M. Izrailev and A.A. Krokhin, \textit{Phys. Rev.
Lett.} \textbf{82}, 4062 (1999).

\bibitem {Tessieri-02}L. Tessieri, \textit{J.Phys. A: Math. Gen.} \textbf{35,}
9585-9600 (2002).

\bibitem {Aspect-private}When this paper was in final preparation, we were
informed about an independent study of the LE in the laser speckle potential,
A. Aspect, private communication and P. Lugan \textit{et al.}, arXiv:0902.0107v2.

\bibitem {Risken}H. Risken, \textit{The Fokker--Planck Equation}, second
edition (Springer-Verlag, Berlin, 1989).

\bibitem {V-Kampen-book}N. G. van Kampen, \textit{Stochastic Processes in
Physics and Chemistry}, revised and enlarged edition (North-Holland,
Amsterdam, 1992).

\bibitem {EGur}paper in preparation

\bibitem {Hangi-95}P. Hanggi and P. Jung, \textit{Adv. Chem. Phys.}
\textbf{89}, 239 (1995).

\bibitem {V-Kampen-89}N. G. van Kampen, \textit{J. Stat. Phys.} \textbf{54},
1289 (1989).

\bibitem {Liu-86}Y. Liu and K.A. Chao, \textit{Phys. Rev. B} \textbf{34}, 5247 (1986).
\end{thebibliography}
\end{document}